\documentclass[prapplied,rsi, amsmath,amssymb,reprint,superscriptaddress]{revtex4-1}

\usepackage{longtable}
\usepackage{graphicx}
\usepackage{color}
\usepackage{soul}

\usepackage[version=3]{mhchem} 
\usepackage{blindtext}
\usepackage{bm}
\usepackage{array}
\usepackage{amsmath}

\newcolumntype{C}{>{\centering\arraybackslash}p{4em}}
\newcolumntype{D}{>{\centering\arraybackslash}p{6em}}

\usepackage{longtable}

\usepackage{array}
\newcolumntype{L}[1]{>{\raggedright\let\newline\\\arraybackslash\hspace{0pt}}m{#1}}
\newcolumntype{C}[1]{>{\centering\let\newline\\\arraybackslash\hspace{0pt}}m{#1}}
\newcolumntype{R}[1]{>{\raggedleft\let\newline\\\arraybackslash\hspace{0pt}}m{#1}}

\begin{document}

\title{Growth driven phase transitions in Zinc Oxide nanoparticles through machine-learning assisted simulations}

\author{Quentin Gromoff}
\affiliation{CEMES, CNRS and Université de Toulouse, 29 rue Jeanne Marvig, 31055 Toulouse Cedex, France}

\author{Magali Benoit}
\affiliation{CEMES, CNRS and Université de Toulouse, 29 rue Jeanne Marvig, 31055 Toulouse Cedex, France}

\author{Jacek Goniakowski}
\affiliation{CNRS, Sorbonne Université, Institut des NanoSciences de Paris, UMR 7588, 4 Place Jussieu, F-75005 Paris, France}

\author{Carlos R. Salazar}
\affiliation{Univ. Lille, CNRS, INRA, ENSCL, UMR 8207, UMET, Unité Matériaux et Transformations, F 59000 Lille, France}

\author{Julien Lam}
\email{julien.lam@cnrs.fr}      
\affiliation{Univ. Lille, CNRS, INRA, ENSCL, UMR 8207, UMET, Unité Matériaux et Transformations, F 59000 Lille, France}

\begin{abstract}
This study investigates the formation of zinc oxide (ZnO) nanoparticles, a material of significant technological interest with complex structural properties, through atom-by-atom deposition modeling—a process common in bottom-up synthesis. Our findings demonstrate that, although the body-centered tetragonal (BCT) structure is thermodynamically stable at equilibrium for small particle sizes, the deposition process induces a crystal-to-crystal phase transition into the more stable wurtzite (WRZ) phase. This transformation is facilitated by a specific redistribution of the nanoparticle ions, which effectively compensates the emerging polar facets at the moment of transition. These insights offer a deeper understanding of oxide nanoparticle formation, which should ultimately help the design of materials with targeted structural features.
\end{abstract}

\maketitle

\section{Introduction}
Since most nanoparticle (NP) applications rely on specific structural properties, the very notion of stability at nanoscale is crucial to understand\,\cite{Phan2019May}. In experiments, studies of nanoparticle stability generally consist in imposing an external constraint like temperature,\cite{Gunawan2008Dec,Gao2015} pressure,\cite{Valencia2018Oct,Carlton2012Nov} reactive gas\cite{Nassereddine2021Nov,Petkov2013Jul} and observing either post-mortem or with in-situ electron microscopy the subsequent structural changes. Meanwhile, from the simulation perspectives, structural stability is usually investigated through energetic considerations. For the smallest NP sizes, density functional theory (DFT) calculations are employed to minimize structures of different morphologies or polymorphic structures and deduce the most stable one by comparing their relative energy \cite{Pizzagalli2024Aug,Vines2017Jul}. For larger NP sizes, DFT is employed to compute cohesive and surface energies. These values are then incorporated into semi-continuum models to estimate the total NP energy and construct NP stability diagrams\,\cite{Gonzalez2014Apr,Barnard2014}. In both cases, contrary to experimental measurements, those studies neglect dynamical restructurings that can occur under more realistic thermodynamic conditions.

From the technical viewpoint, the main bottlenecks that hinder our ability to dynamically observe phase transitions in nanoparticles are the large computational cost of DFT calculations and the lack of accurate interatomic force fields. Recently, machine-learning interatomic potentials (MLIP) have demonstrated their ability to retain the accuracy of DFT calculations while reaching computational costs comparable to classical force fields\,\cite{Nyangiwe2025Jul}. 

Building on the recent advances in MLIPs, this work focuses on ZnO NPs, as they combine technological interests in electrochemistry\,\cite{Zhou2023Apr,Nagajyothi2013Oct,Sun2016Nov} and bacteriology\,\cite{Matinise2017Jun,Pushpalatha2022,Islam2022Mar,Gudkov2021Mar} with fundamental challenges. Indeed, DFT calculations have predicted a size-dependent crossover between body-centered tetragonal (BCT) and wurtzite (WRZ) structures at 0 K.\cite{Vines2017Jul} More recently, employing the latest version of the Physical LassoLars Interaction Potentials including long-range interactions (PLIP+Q), we investigated crystal nucleation dynamics starting from a liquid nanodroplet.\cite{Salazar2024Aug} Our results revealed distinct nucleation pathways depending on the degree of supercooling, as determined by the system temperature. In all cases, the system predominantly crystallized into WRZ structures; however, under deep supercooling, BCT nuclei appeared transiently at the early stages, whereas WRZ emerged directly at higher temperatures. Importantly, the explicit treatment of long-range interactions proved essential for accurately capturing surface polarity effects, which remain inaccessible to traditional MLIPs.

In this article, we used the same force field to simulate atom-by-atom deposition on ZnO NPs. In particular, simulations were initialized with  cluster seeds of different sizes and crystal structures. We then employed a  data-driven methodology in the form of the Steinhardt Gaussian mixture analysis (SGMA) to identify the obtained crystalline ordering \cite{Furstoss2025Apr}. For both initial crystal structure seed, we showed that WRZ usually emerges at the end of the simulations. For the largest sizes, this can naturally be explained by the more stable bulk cohesive energy of WRZ. We show that the BCT-to-WRZ transition is consistently accompanied by a compensation in the polarity of the emerging WRZ basal facets and that the substantial ionic separation required for this process is achieved very efficiently under the growth conditions.

These findings offer new insights into the inherent complexity of nanoparticle formation with a particular focus on oxides which have been far less studied. Crucially, they also highlight both the necessity of incorporating long-range interactions and PLIP+Q's ability to do so accurately within MLIPs.

\section{Methodology}

\subsection{Charge included Physical Lassolars Interaction Potential (PLIP+Q)}

To model ZnO NPs, we use a previously obtained MLIP\,\cite{Salazar2024Aug}. Herein, we will only briefly summarize some of the most important features of our implementation. 

The PLIP+Q model can be split in a short-range and a long-range interaction. For the short-range part, the atomic contributions $E_i$, for each atom  $i$ of the system is written as a linear combination of 2-body, 3-body, and N-body descriptors $\{\chi\}$.

\begin{equation}
E^{i}_{\text{short}} = \sum_{n} \omega_n \chi_n^{i}
\end{equation}

\begin{equation}
\chi_n^{\text{2B},i} = \sum_{j} f_n(r_{ij}) f_c(r_{ij})
\end{equation}

\begin{equation}
\chi_{n,l}^{\text{3B},i} = \sum_{j} \sum_{k} f_n(r_{ij}) f_c(r_{ij}) f_c(r_{ik}) \cos^l(\theta_{ijk})
\end{equation}

\begin{equation}
\chi_{n,m}^{\text{NB},i} = \left[ \sum_{j} f_n(r_{ij}) f_c(r_{ij}) f_j(r_{ij}) \right]^m
\end{equation}

with $r_{ij}$ the distance between the atoms $i$ and $j$, $\theta_{ijk}$ the angle around the atom $i$, $\omega_n$ the fitted linear coefficients, $f_n$ Gaussian functions, $f_c$ a cutoff function, $f_s$ a polynomial function, and $l$ and $m$ two positive integers with $l\leq5$ and $m\leq7$. In particular, for the Gaussian functions, we used central positions $R_s$ ranging from 0.5 to 6.0 and widths $\eta$ ranging from 0.5 to 1.5 with steps of 0.5 in both cases. Altogether, it leads to a total of 1980 descriptors. The $\omega_n$ coefficients  are obtained during the training process. The cutoff radius of the short-range potential is fixed at 6\,\AA. For the fitting procedure, we used the LassoLars regression which allows us to select the most valuable descriptors thus reducing the complexity of the model and achieve better computational efficiency.

In order to better model the complexity related to polar surfaces in ZnO, we complete the short-range PLIP with long-range coulombic interactions. In PLIP+Q, static point charges are used to describe electrostatic interactions. These charges remain fixed with time and the local environment, and are initially set as the oxidation numbers of the atomic species. The fictive electrostatic contribution to the energy $E^i_{el}$ is computed using the point charge Coulomb model and the Ewald summation method, and it's effective contribution is weighted by the charge scaling coefficient $\gamma$.
\begin{equation}
    E^i = E_{\text{short}}^i + \gamma E_{\text{el}}^i
\end{equation}
In practice, the $\gamma$ parameter is fitted along with the $\omega_n$ coefficients of the short-range contribution during the LassoLars training. 

For the training database, crystal, surface, and liquid configurations were used for a total of 70471 atomic environments. The crystal and surface structures belong to the 7 different ZnO polymorphs and are sampled around their equilibrium position using MD simulations using a Buckingham potential to begin and sequentially better PLIP models\,\cite{Goniakowski2022Oct}. For each snapshot, the forces were calculated within the framework of density functional theory (DFT) using the Vienna Ab initio Simulation Package (VASP) with the Perdew-Wang exchange correlation functional (PW91)\,\cite{Kresse1993,Kresse1996}. The Projector Augmented Wave (PAW) method was employed \cite{Kresse1999}, employing a soft s2p4 oxygen pseudopotential and a d10p2 zinc pseudopotential for the valence electrons. Further details on the mathematical framework, the employed database, and the validation procedure are available in our previous publications.\cite{Goniakowski2022Oct,Salazar2024Aug} 

The PLIP+Q model was previously tested for many physical properties including lattice spacing, polymorph stability, radial distribution functions, surface energies and nanoparticle stability\,\cite{Salazar2024Aug}. In addition, to specifically verify our usage in growth conditions, we performed additional single point calculations [See SI A].

\subsection{Growth Simulations} \label{Growth}

\begin{figure}[ht]
    \centering
    \includegraphics[width = 7cm]{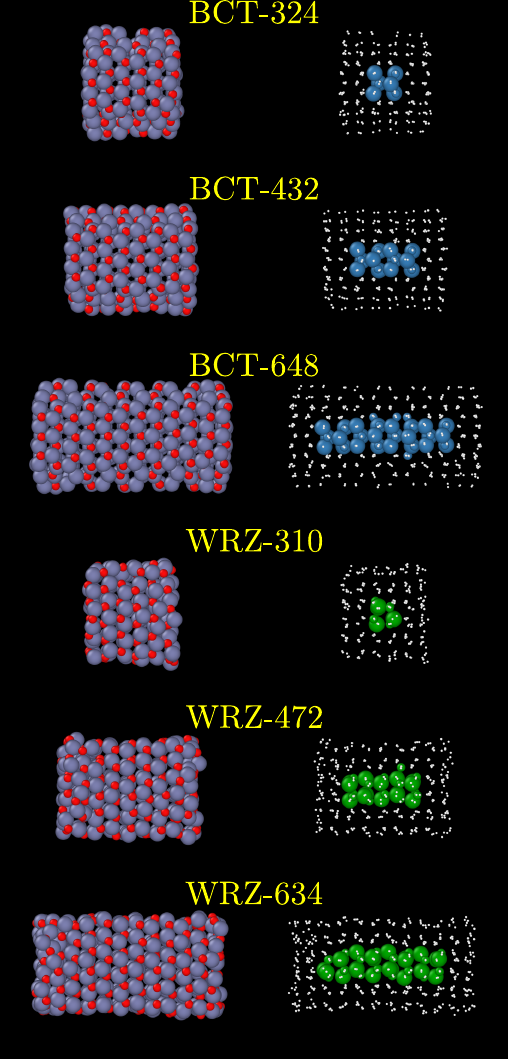}  
    \caption{Left: atomic structures of all the initial seeds considered, in the BCT or WRZ structural arrangements (gray: Zn, red: O). Right: corresponding analysis of the local atomic order using the SGMA analysis (blue: BCT, green: WRZ).}
  \label{fig:InitialSeeds}
\end{figure}

Our growth simulations are initialized with different ZnO clusters of few hundreds of atoms, named hereafter "initial seeds" [see Fig.\protect\ref{fig:InitialSeeds}]. Those clusters are of two types depending on whether their crystalline structure is body-centered tetragonal (BCT) or wurtzite (WRZ). They were obtained by cutting finite clusters from the corresponding bulk phases. Specifically, the initial WRZ seeds were modeled as elongated hexagonal prisms bounded by opposing (0001)/(000–1) basal facets and (11–20)/(10–10) side facets. They contain 310, 472, and 634 atoms, corresponding to 6, 9, and 12 layers of 27 ZnO formula units. Because the (0001) and (000–1) surfaces are Zn- and O-terminated, the polarity was compensated by surface reconstruction: 7 of 27 terminal cations and anions were removed from the Zn- and O-terminated basal facets, respectively. It is important to emphasize that an accurate assessment of the stability of polar versus non-polar terminations demands explicit treatment of long-range electrostatics. Indeed, we have previously demonstrated that a short-range potential alone underestimates the energy of uncompensated polar terminations, incorrectly predicting them to be more stable than non-polar facets\,\cite{Salazar2024Aug}. Then, in the same paper, we included long-range electrostatics which allowed us to drastically reduce this error and to restore the correct DFT hierarchy between non-polar, compensated polar, and uncompensated polar surfaces. Analogous BCT seeds were constructed in the same hexagonal prism shape, terminated by opposing (100) facets and (010)/(011) side ones. These comprise 324, 432, and 648 atoms, corresponding to 6, 8, and 12 layers of 27 ZnO formula units. Unlike WRZ, all BCT facets are non-polar and required no ionic excess or deficiency. For each of these initial seeds, the atomic positions were first relaxed using the PLIP+Q potential to find the energy minimum and then further equilibrated at a given temperature for 10\,ps.

For the subsequent deposition, a spherical shell with a radius of 25\AA\ is defined around the nanoparticle, within which a zinc atom and an oxygen atom are introduced at fixed time intervals and at random positions. The newly introduced atoms are assigned initial velocities directed toward the center of mass of the system in order to enforce deposition and growth. Atoms are deposited with rate of one pair of Zinc and Oxygen atoms every 10 picoseconds which is by far slower than with ab initio MD while remaining several orders of magnitudes faster than experiments\,\cite{Zhao2016Apr}. It is to be noted that the chosen deposition rate is faster than what can typically be achieved with classical force fields. This is due to the computational cost of the PLIP+Q model, which requires 20 hours to produce 1 ns of simulation time using 32 CPUs under our growth conditions. Meanwhile, we also tested deposition rate that were five times slower and obtained qualitatively similar results [See SI B]. Between two depositions, a molecular dynamics (MD) simulation of the system is carried out at a given temperature in the NVT ensemble. We note that the stochastic nature of the process leads to some depositions being unsuccessful ie. the deposited atoms bounce on the surface and return in the gas phase. Those particular atoms are removed from the simulation box and the the growth simulation is terminated when 200 ZnO depositions were attempted. For each initial cluster and thermodynamic condition, 10 different growth simulations are performed starting from  different initial conditions for the velocities and the random positions of the deposited atoms. In addition, we also measured the surface mobility of the atoms and verified that our conditions lead to enough mobility for atomic reconstruction [See SI C]. The trajectories are generated using the Large-scale Atomic/Molecular Massively Parallel Simulator (LAMMPS).\cite{Thompson2022Feb} All the simulations are conducted with a timestep of 1\,fs in the NVT ensemble  with a temperature controlled by the Nosé–Hoover thermostat using a damping parameter of  100\,fs.

\subsection{Local Ordering Analysis} \label{SGMA}

Characterization of the local ordering in ZnO NPs requires a method capable of discriminating between the multiple ZnO polymorphs. For that purpose, we use the Steinhardt Gaussian mixture analysis (SGMA) \cite{Salazar2024Aug,Furstoss2025Apr} that we developped previously and tested on ZnO nanoparticle simulations but also on \ce{Mg2SiO4} defects and \ce{Al2O3} nanoparticles.

First, a database made of different configurations is obtained by running NVT simulations on the various crystals with a temperature ramp from 200\,K to 1500\,K. Additional liquid structures at 2500\,K are also included, resulting in a total of 21 snapshots per crystals and 21 for liquid structures. Then, for each configuration, the averaged Voronoi-weighted Steinhardt parameters $q_2$ to $q_8$ are computed. This step is performed twice: first using the full list of atomic neighbors, and second, considering only atoms of the same species, yielding a total of 14 order parameters for each atom. These order parameters are then used to construct the input data for training. Next, the classification part is done by a Gaussian Mixture Model (GMM), that is trained to try to represent each structure type by a single Gaussian distribution in the space defined by the order parameters. This way, it is later possible to map a Gaussian cluster to the matching structure by comparing with the descriptors of the perfect configurations. The training process is handled by the Expectation-Maximization algorithm that estimates the optimal parameters for the Gaussian components through multiple iterations. Each Gaussian component is defined by a mean vector and covariance matrix, with mixture weights reflecting the relative probability of an atom belonging to one of the 8 structure types in the database. Finally, proper classification is achieved with a Maximum Likelihood approach: atoms are assigned to the Gaussian cluster yielding the highest probability. Moreover, the probability needs to be above 50\% as an additional criterion to improve reliability. For more details on the methodology and its implementation, please refer to our previous publications.\cite{Salazar2024Aug,Furstoss2025Apr}

In the original publication,\cite{Salazar2024Aug} this classification framework was validated on nanoparticles seeded with crystalline phases. Here, we further tested the model for each initial seed [see Fig.\,\ref{fig:InitialSeeds}]. The blue and green atoms represent the atoms that are identified as BCT and WRZ atoms, respectively. Since the  atomic environment of the surface and subsurface atoms is not equivalent to that of the bulk, these atoms are not recognized by the SGMA method as belonging to a crystal structure of the database. They therefore appear in  white in Fig.\,\ref{fig:InitialSeeds}.


\section{Results}

\subsection{Motivation and preliminary calculations}

\begin{figure}[ht]
    \centering
    \includegraphics[width = 8.6cm]{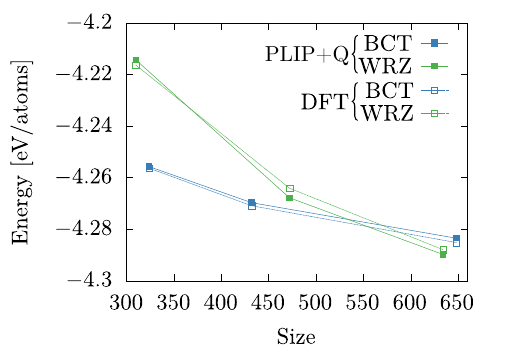}  
    \caption{Energy per atom of NPs in the BCT and WRZ phase as a function of size (number of atoms) computed with DFT and PLIP+Q.}
\label{fig:energies}
\end{figure}

To contextualize the following growth simulations, we will first explore some equilibrium properties of ZnO at the nanoscale. In particular, the six initial seeds shown in Fig.\,\ref{fig:InitialSeeds} were relaxed with the PLIP+Q potential and with DFT using the framework employed for the MLIP database. The final energies per atom are reported in Fig.\,\ref{fig:energies} as a function of NP size. Both calculations show the same trend: BCT-type NPs are more stable than WRZ-type NPs for small sizes, and a reverse stability is observed around 550 atoms for PLIP+Q and 580 atoms for DFT. 

The enhanced stability of BCT-type nanoparticles at small sizes arises from their more favorable surface energy, which compensates for the more stable bulk of the WRZ phase\,\cite{Sponza2015Feb,Sponza2016May}. Then, beyond a certain size, the gain in bulk energy, proportional to the number of atoms, outweighs the gain in surface energy thus reestablishing the stability of WRZ. We note that the crossover size obtained here is smaller than that reported by Vines et al.\cite{Vines2017Jul}, as our WRZ clusters were stabilized by compensating the polarity of their basal facets (see Sec. \ref{Growth}).

In order to verify whether the relative stability of the two types of NPs is maintained with temperature, NVT molecular dynamics simulations were performed at 500\,K, 700\,K, and 900\,K for each of the initial clusters during 2\,ns. The number of atoms of each crystalline type—BCT or WRZ—inside the core was monitored during the MD trajectories. In all cases, the NPs remain in their initial crystalline phase whatever the temperature. In particular, we did not observe a transition from the BCT to the WRZ type for the largest NP, implying that the underlying free energy barrier is quite high. 

\vspace{0.25cm}
From the experimental viewpoint, in bottom-up synthesis—including wet chemistry, gas-phase combustion, and ablation methods—nanoparticles are formed by the successive addition of atoms that aggregate into small cluster seeds. In the ZnO case, due to competition at small sizes, one may wonder whether a phase transition favoring WRZ can occur when starting with a small BCT-type cluster or if the system can become kinetically trapped in the BCT phase. This goes beyond applicative purposes and raises fundamental questions regarding solid-solid transitions at the nanoscale\,\cite{Machon2013Dec} and polymorph selection in the presence of a gaseous interface.\cite{Schoonen2022Dec} To study this challenging system, atom-by-atom deposition simulations were performed starting with different initial cluster seeds.

\subsection{General behavior}

\begin{figure}[ht]
    \centering
    \includegraphics[width = 8.6cm]{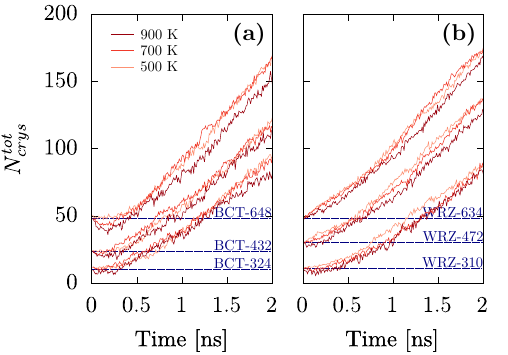}
    \caption{Evolution of the number of atoms in the crystalline nucleus as a function of time, averaged over the 10 independent simulations for (a) BCT-324, BCT-432 and BCT-648 initial seeds and (b) WRZ-310, WRZ-472 and WRZ-634 initial seeds, at 500\,K, 700\,K and 900\,K.}
    \label{fig:ncrys-time}
\end{figure}

Ten independent atom-by-atom deposition simulations were performed starting from the BCT and WRZ seeds of Fig.\,\protect\ref{fig:InitialSeeds} at three different temperatures: 500\,K, 700\,K, and 900\,K. In all cases, using SGMA, we observed a linear increase in the total number of atoms within the crystalline core, $N_{crys}^{tot}$, as shown in Fig.\,\ref{fig:ncrys-time}.

For the BCT seeds (Figure \protect\ref{fig:ncrys-time}(a)), the number of atoms in the crystalline nucleus initially decreases slightly before growing linearly. The growth rate is somewhat higher at 500\,K than at 900\,K (Table \protect\ref{tab:growth-rates}), which may be attributed to increased structural disorder at elevated temperatures, potentially hindering the crystallization process.

In contrast, for WRZ initial seeds, the crystalline core begins growing almost immediately after deposition starts, with a slight initial delay only in the smallest clusters. As anticipated, larger initial clusters exhibit higher growth rates than smaller ones. However, the growth rate of WRZ seeds appears to be independent of temperature, as indicated in Table\,\protect\ref{tab:growth-rates}.

\begin{table}[ht]
\caption{Growth rates in atom/ns averaged over the 10 independent simulations for the two crystalline types, the three initial seed sizes and the three temperatures.}
\label{tab:growth-rates}
\begin{tabular}{l|ccc}
\hline
  & 500\,K & 700\,K & 900\,K \\
  \hline
BCT-324 & 56 & 51 & 42 \\
BCT-432 & 71 & 56 & 51 \\
BCT-648 & 80 & 78 & 70 \\
\hline 
WRZ-310 & 47 & 59 & 56 \\
WRZ-472 & 59 & 71 & 66 \\
WRZ-634 & 76 & 64 & 75 \\
\hline
\end{tabular}
\end{table}

\begin{figure}[ht]
    \centering
    \includegraphics[width = 9cm]{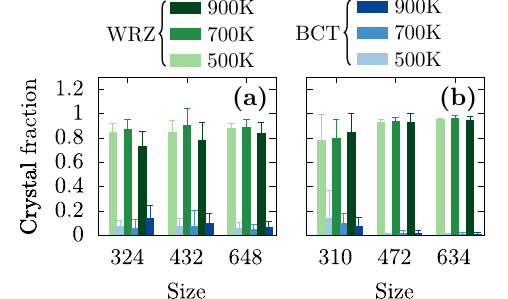}
    \caption{Fraction of atoms of the WRZ (green) and BCT (blue) types inside the crystalline nucleus at the end of the simulations, averaged over the 10 independent simulations, for (a) the BCT initial seeds and (b) the WRZ initial seeds at 500K, 700K and 900K. }
    \label{fig:EndPoint}
\end{figure}

Fig.\,\ref{fig:EndPoint} shows the final fraction of atoms in each structural phase within the crystalline core, averaged over 10 independent simulations, for all initial seeds and temperatures. In most cases, the WRZ phase accounts for more than 80\% of the core. For BCT-initialized seeds at 900 K, the WRZ fraction falls slightly below 80\% for the smallest clusters, which can be attributed to rare instances where the structural transition did not occur or the core remained poorly crystallized. A similar reduction in WRZ fraction is observed for the smallest WRZ initial clusters. In that case, the decrease arises from a few simulations where the crystalline nucleus reverted to the BCT phase, as will be discussed in more detail below. 

As such, Fig.\,\ref{fig:EndPoint} clearly demonstrates that the BCT-to-WRZ structural transition occurs in the vast majority of deposition simulations, irrespective of the initial cluster size or temperature. This behavior stands in stark contrast to the results obtained from simulations performed at fixed particle size initialized with crystalline seeds, where no transition from BCT to WRZ was observed, even for the largest nanoparticles.

\begin{figure}[ht]
\centering
    \includegraphics[width = 8.6cm]{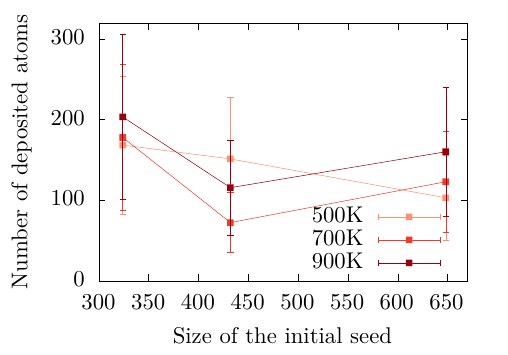}
    \caption{Number of deposited atoms at the BCT-to-WRZ transition as a function of the initial BCT seed at 500K, 700K and 900K, averaged over the 10 independent simulations.}
    \label{fig:Transition}
\end{figure}

To assess whether the nanoparticle (NP) size is the key parameter governing the BCT-to-WRZ structural transition, the number of deposited atoms at the transition point has been plotted as a function of the initial BCT seed size in Fig. \protect\ref{fig:Transition}, with data averaged over 10 independent simulation simulations. This transition point is defined as the last moment when the number of atoms in the BCT-like configuration surpasses those in the WRZ-like configuration. As previously noted, temperature appears to have a minimal influence on the transition behavior. More surprisingly, it seems that the number of deposited atoms at the transition point does not really depends on the size of the seed as well. We note that the measured error bars are relatively large. But, using a convergence of the mean analysis, we can show that 10 independent simulations is enough [see SI D] which means that the large error bars are only due to the stochastic nature of the crystallization process.

\begin{figure}
    \centering
    \includegraphics[width = 8.6cm]{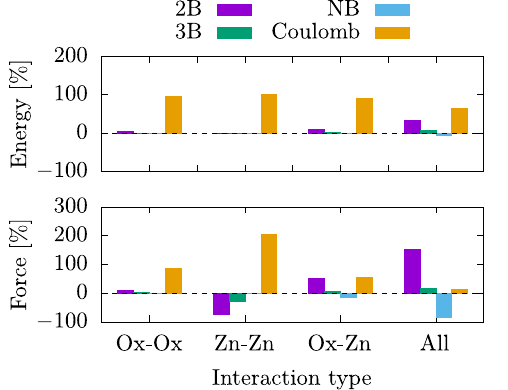}
    \caption{Contributions of 2-body, 3-body, N-body, and electrostatic terms to the interaction energies and interatomic forces for oxygen–oxygen, zinc–zinc, and oxygen–zinc pairs, obtained by averaging over all atoms and the entire growth simulation time.}
    \label{Rationalize}
\end{figure}

Before detailing the microscopic process occurring during the growth, we would like to return to the PLIP+Q usage. In particular, a primary advantage of PLIP+Q over conventional MLIP models is its explicit decomposition of interactions into 2-body, 3-body, N-body, and electrostatic components. We leveraged this framework to gain physical insights into the nature of atomic interactions in zinc oxide. By analyzing our growth simulations, we calculated the contribution of each term — averaging the atomic energy and forces across all atoms in all of the 75 NP configurations considered during the validation of the PLIP+Q models done in SI\,A (see Fig.\,\ref{Rationalize}). Our analysis revealed several key findings. Firstly, not all components contribute with the same sign to the total interaction. Terms are denoted as positive when they align with the sign of the total interaction and negative when they oppose it. Secondly, energy and forces are not partitioned equivalently. Some interaction terms contribute significantly to the total energy while remaining near a local minimum for the corresponding gradients, resulting in minimal force contributions. Thirdly, comparing the four terms shows that Coulombic interactions are generally the most dominant in the energy landscape. Yet, when summing homonuclear and heteronuclear contributions, the Coulombic forces largely cancel out [See Fig\,\ref{Rationalize}.(b) "All"  contributions]. Consequently, 2-body and N-body terms become the primary drivers of the total force. This cancellation likely stems from the use of uniform electrostatic charges for the Zn and O atoms.

\subsection{Microscopic picture of the growth mediated phase transition}

\begin{figure*}[ht]
    \centering
    \includegraphics[width = 1.\linewidth]{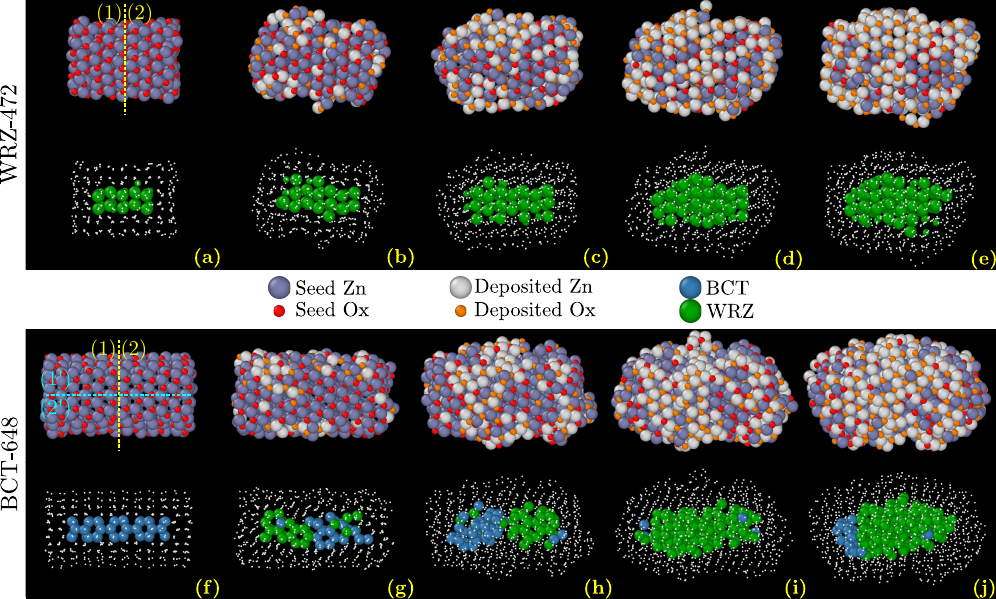}
    \caption{Selected snapshots of the nanoparticle during the deposition simulations. (a) to (e): WRZ-472 initial seed. (f) and (j): BCT-648 initial seed. In each case, the top images show the entire NP with the deposited atoms in different color codes. The bottom images show the crystalline cores with the BCT atoms in blue and the WRZ atoms in green. In Fig.\,\ref{fig:Frise}(a,f), the dashed lines correspond to the planes used for computing the Zn excess.}
    \label{fig:Frise}
\end{figure*}

To further illustrate the dynamics of the transition, Fig. \protect\ref{fig:Frise} presents the temporal evolution of the crystalline core structure for simulations starting from a WRZ-472 seed and a BCT-648 seed. In the top sequence of images, the evolution of the crystalline core is shown for a WRZ initial seed. In this first case, the WRZ phase grows continuously and remains dominant throughout the deposition process. In the bottom sequence of Fig. \protect\ref{fig:Frise} [Figs. \protect\ref{fig:Frise}(f)–(j)], the initially BCT-type crystalline core gradually transforms into the WRZ phase [Figs. \protect\ref{fig:Frise}(b) and (c)]. During an intermediate stage, the two phases coexist within the core before the WRZ structure progressively dominates. This confirms that the transition is not abrupt but rather occurs through a gradual phase replacement process. It is also worth noting that in both cases—regardless of the initial structure—the crystalline core maintains a well-defined order, even though the outer surface of the NP appears rough and disordered due to ongoing atomic deposition.

\begin{figure*}[ht]
    \centering
    \includegraphics[width = 1.\linewidth]{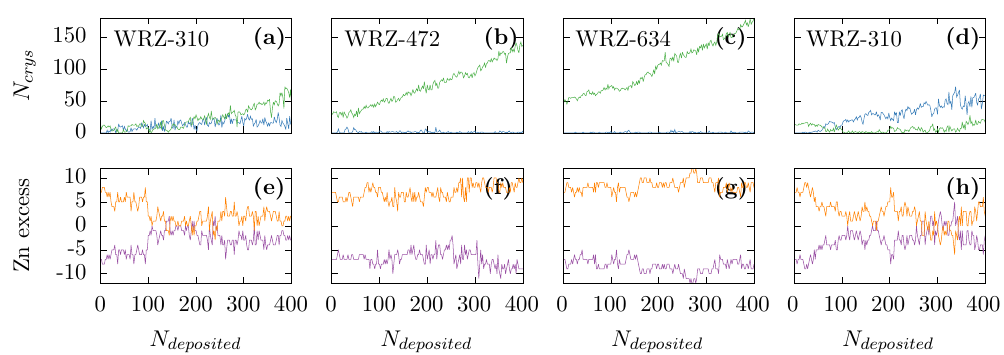}
    \caption{((a) to (c): Examples of the evolution of the number of atoms of the BCT (blue) and WRZ (green) type inside the crystalline nucleus as a function of the number of deposited atoms for different WRZ initial seeds at 700\,K. (e) to (h): Excess of Zn atoms at opposite ends of the nanoparticle, in orange and purple, respectively. (d) and (h): same at 500\,K.}
    \label{fig:VsTime_WRZ}
\end{figure*}

\begin{figure*}[ht]
    \centering
    \includegraphics[width = 1.\linewidth]{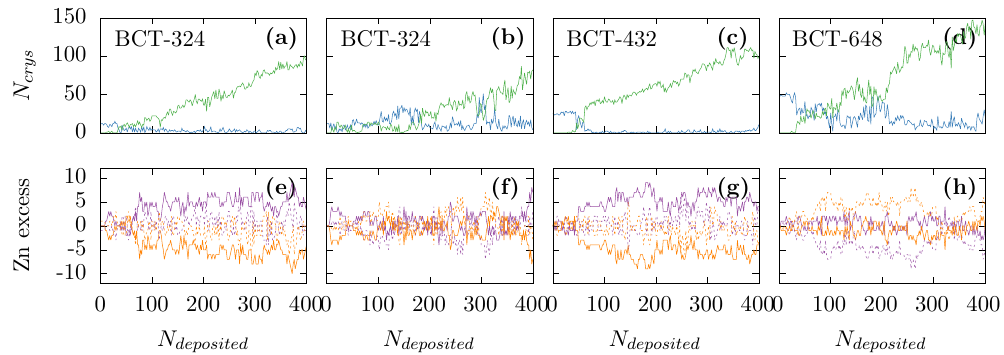}
    \caption{((a) to (d): Examples of the evolution of the number of atoms of the BCT (blue) and WRZ (green) type inside the crystalline nucleus as a function of the number of deposited atoms for different BCT initial seeds at 700\,K. (e) to (h): Excess of Zn atoms at opposite ends of the nanoparticle. In dashed lines, the Zn excess in the perpendicular direction. }
    \label{fig:VsTime_BCT}
\end{figure*}

To gain deeper insight into the mechanisms governing the BCT-to-WRZ transition, it is instructive to first recall the distinct nature of their terminal facets. The opposing (100) end facets of BCT-type nanoparticles are stoichiometric and therefore non-polar. In contrast, the (0001) and (000–1) end facets of WRZ-type nanoparticles are polar and thus require an excess of O and Zn atoms, respectively, to achieve electrostatic stabilization.\cite{Goniakowski2007Dec} Consequently, the BCT-to-WRZ transformation inherently involves the development of ionic separation, which manifests as opposite compensating ionic excesses at the opposing NP's facets. To monitor this effect during deposition simulations, the charge imbalance was systematically quantified at both terminations. For that purpose, we first remove any stochastic translation and rotation of the system. We then split the NPs into two equal halves by a plane passing through the center of the nanoparticle and oriented parallel to the original basal facets [See Fig.\,\ref{fig:Frise}(a,f)]. For each half, we calculate the difference between the number of zinc and oxygen atoms, which we term Zn excess. In few of the BCT cases, growth can also occur perpendicularly which we also considered by using a perpendicular plane [See Fig.\,\ref{fig:Frise}(f)]. Figures \protect\ref{fig:VsTime_WRZ} and \protect\ref{fig:VsTime_BCT} illustrate representative examples of this analysis, displaying the temporal evolution of the number of atoms associated with each crystalline phase within the nucleus, along with the corresponding variation of Zn excess for WRZ and BCT initial seeds, respectively.

Focusing first on the growth of nanoparticles from WRZ seeds — where, as described in the Methods section, the required ionic excess was incorporated into the initial configurations — we find that particles initiated from the larger seeds (Figs. \protect\ref{fig:VsTime_WRZ}(b, c)) exhibit steady growth of their WRZ crystalline cores at comparable rates ($\Delta N_{crys}/\Delta N_{deposited} \sim$ 0.25). In both cases, the initial Zn excess is largely preserved throughout deposition, indicating that the added atoms primarily contribute to elongation of the crystalline core and development of a less ordered peripheral zone, with only minor impact on core width. This preservation of Zn excess also suggests the absence of additional polarity effects induced by the small particle size\,\cite{GoniakowskiPRB2011,NogueraCR2013}. In contrast, in smaller seeds (Figs. \protect\ref{fig:VsTime_WRZ}(a, d)), the initial WRZ phase disappears nearly completely due to its thermodynamic instability at the smallest particle sizes. Its disappearance is followed by a rapid decrease of the initial Zn excess, thus demonstrating that ionic mobility during deposition efficiently modulates local NP composition in response to changes in the character of the exposed facets. Partial recovery of the WRZ phase is observed at later growth stages (Fig. \protect\ref{fig:VsTime_WRZ}(a)), although this recovery can be significantly delayed by the presence of a well-developed BCT core (Fig. \protect\ref{fig:VsTime_WRZ}(d)). In both these cases, the overall WRZ fraction remains low, resulting in only marginal ionic separation. We note that the process shown in Fig. 7(a) and Fig. 7(d) occurs only five times across our 90 simulations. We observe that, in all cases where growth of WRZ phase is substantially delayed, the Zn excess is not being maintained  during the atom-by-atom deposition [See SI E]. In contrast, as we will show next, emergence of a Zn excess can actually induce a phase transition when starting from BCT seeds.

Turning to nanoparticle growth from BCT seeds (Fig. \protect\ref{fig:VsTime_BCT}), we find that the progressive emergence of the WRZ phase — observed consistently across all cases — correlates with markedly different behaviors in Zn excess.In cases \protect\ref{fig:VsTime_BCT}(a and c), where the BCT-to-WRZ transformation occurs at an early stage of deposition, the steady growth of the WRZ crystalline core closely matches that of larger WRZ-seeded particles (Figs. \protect\ref{fig:VsTime_WRZ}(b and c)), both in growth rate ($\Delta N_{crys}/\Delta N_{deposited} \sim$ 0.25) and in the complete absence of residual BCT content. These systems also exhibit rapid establishment of stable ionic separation, with Zn excess and deficiency localized at opposite nanoparticle ends, consistent with the polarity of the two basal facets of the growing WRZ core. In contrast, in case \protect\ref{fig:VsTime_BCT}(b), the formation and growth of the WRZ core are substantially delayed by the persistence of well-defined BCT grains. Steady WRZ growth commences only near the end of the simulation and is accompanied by a sharp onset of ionic separation. This behavior closely resembles that of small WRZ-seeded nanoparticles (Figs. \protect\ref{fig:VsTime_WRZ}(a and d)), confirming that for the smallest seeds their initial structural configuration may be of limited importance since the high structural flexibility at the considered temperature allows thorough reorganization during growth.  Finally, case \protect\ref{fig:VsTime_BCT}(d) presents a less straightforward scenario. Here, competition between the initial BCT phase and the emerging WRZ phase results in a delayed and overall less steady increase of WRZ content, with residual BCT persisting throughout the entire simulation. In this case, the polar direction of the WRZ phase  still emerges but in the perpendicular direction with respect to the initial NP long axis [See dashed lines in Fig\,\ref{fig:VsTime_BCT}(h)].

\begin{figure}[h!]
    \centering
    \includegraphics[width = 8.6cm]{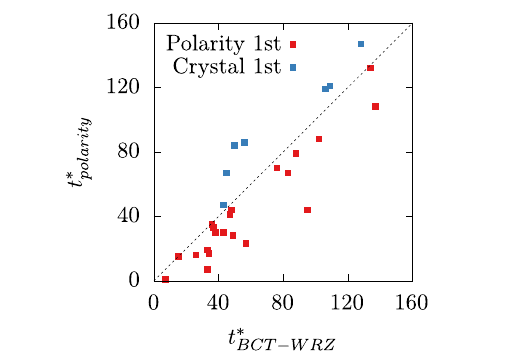}
    \caption{Correlation between polymorphic transformation ($t_{BCT \rightarrow WRZ}^*$ is the last moment when  $N_{BCT}>N_{WRZ}$) and  ionic redistribution ($t_{excess}^*$ is the last moments when $\mid Zn_{excess}^{(1)}-Zn_{excess}^{(2)} \mid<2$) during BCT$\rightarrow$WRZ transition in selected simulations initiated from BCT seeds of varying sizes and at different temperatures. }
    \label{WhoFirst}
\end{figure}

To make our analysis more quantitative, we focused on simulations initiated from BCT seeds of varying sizes and at different temperatures, discarded ambiguous cases exhibiting coexistence of BCT and WRZ structures, and retained only simulations showing clear WRZ growth and a well-established ionic imbalance at the end. From this selected set, we computed two characteristic times: (1) $t_{BCT \rightarrow WRZ}^*$ which corresponds to the last moment when $N_{BCT}>N_{WRZ}$ and (2) $t_{excess}^*$ which corresponds to the last moment when $\mid Zn_{excess}^{(1)}-Zn_{excess}^{(2)} \mid<2$. Fig.\,\ref{WhoFirst} shows that the ionic redistribution tends to precede the polymorphic transformation, especially if the BCT$\rightarrow$WRZ transition occurs rapidly ($t_{BCT \rightarrow WRZ}^*<0.5$\,ns)

\section{Conclusions}

In a context where designing NPs with targeted structural features through specific synthesis conditions has become crucial for many applications, our work provides general insights on oxide nanoparticle formation. We specifically concentrate on zinc oxide which exhibits technological interest and offers intriguing fundamental challenges regarding its structure. Here, we modeled atom by atom deposition processes as they occur in most bottom-up synthesis experiments.

We demonstrated that, although the less thermodynamically favored BCT structure remains stable in small particles under equilibrium conditions, the atom-by-atom deposition process induces a crystal-to-crystal transition toward the more stable WRZ phase. This transition is facilitated by a redistribution of surface ions, which enables the compensation of emerging polar facets precisely at the point of transformation.

More generally, while nanoscale polymorphic transitions are generally interpreted as a competition between surface and volume energy contributions, our work highlights an additional level of complexity arising from growth-induced polarity imbalance, which may also occur during the formation of other metallic oxides including titanium, iron and copper\,\cite{Goniakowski2007Dec}. In a context where the role of long-range interactions is being considered by MLIP developers with increasingly more complex approaches, \cite{Behler2021Jul,Ko2023Jun,Gao2022Mar,Anstine2023Mar}, using the simple PLIP+Q model was pivotal for the success of this study as it enables accurate modeling of both polar and non-polar surfaces at reasonable computational cost through static fixed-point charge electrostatics.

\section*{Acknowledgments}
We acknowledge funding from the French government through the French National Research Agency (ANR) (ANR-21-CE09-0006). Computational resources have been provided by the DSI at Université de Lille, CALMIP, Jean-Zay and TGCC.

\end{document}


\title{Growth driven phase transitions in Zinc Oxide nanoparticles through machine-learning assisted simulations}

\author{Quentin Gromoff}
\affiliation{CEMES, CNRS and Université de Toulouse, 29 rue Jeanne Marvig, 31055 Toulouse Cedex, France}

\author{Magali Benoit}
\affiliation{CEMES, CNRS and Université de Toulouse, 29 rue Jeanne Marvig, 31055 Toulouse Cedex, France}

\author{Jacek Goniakowski}
\affiliation{CNRS, Sorbonne Université, Institut des NanoSciences de Paris, UMR 7588, 4 Place Jussieu, F-75005 Paris, France}

\author{Carlos R. Salazar}
\affiliation{Univ. Lille, CNRS, INRA, ENSCL, UMR 8207, UMET, Unité Matériaux et Transformations, F 59000 Lille, France}

\author{Julien Lam}
\email{julien.lam@cnrs.fr}      
\affiliation{Univ. Lille, CNRS, INRA, ENSCL, UMR 8207, UMET, Unité Matériaux et Transformations, F 59000 Lille, France}


\maketitle

\section{PLIP+Q validation on growth conditions}

\begin{figure}[h!]
    \centering
    \includegraphics[width=17cm]{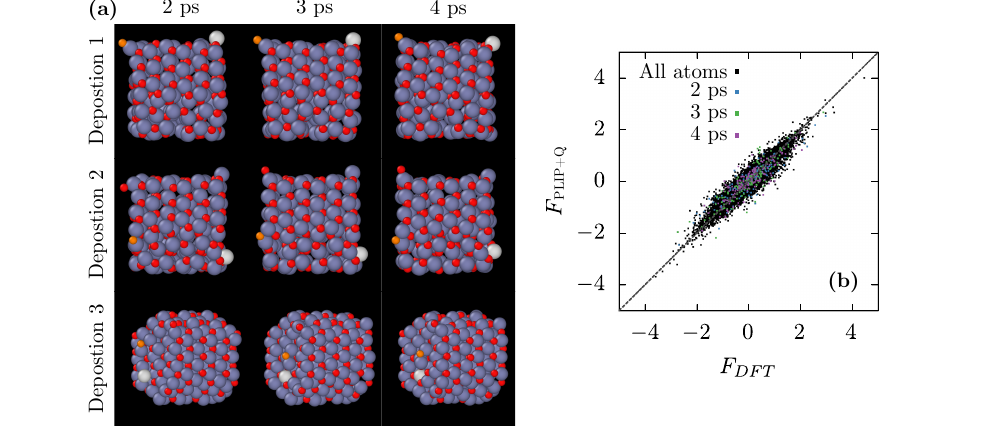}
    \caption{(a) Typical atomic configurations at 2 ps, 3 ps and 4 ps after the successive depositions of Zn and O atoms in the BCT-324 case. We used grey, red, white and oranges to color respectively original Zn and Ox atoms and deposited Zn and Ox atoms. (b) Comparison of atomic forces obtained from DFT with those predicted by PLIP+Q. Forces on just deposited atoms are highlighted in color, while touse on atoms already present in the nanoparticle are shown in black. }
    \label{PLIPvsDFT}
\end{figure}

In order to assess the reliability of the PLIP+Q model under growth conditions, we performed single point DFT calculations using structures extracted from the PLIP+Q MD simulations. Specifically, we focused on the WRZ-310 and BCT-324 cases and selected configurations at 2 ps, 3 ps and 4 ps after the successive depositions of Zn and O atoms [See Fig.\,\ref{PLIPvsDFT}.a]. In total, we have considered 75 configurations spanning different deposition positions. Figure \ref{PLIPvsDFT}.b compares the atomic forces obtained from DFT with those predicted by PLIP+Q. The just deposited atoms are highlighted in color, while atoms already present in the nanoparticle are shown in black. The plot reveals no significant difference between the two atom sets, indicating a consistent force prediction by PLIP+Q, regardless of atom origin. The computed root-mean-square error is 0.22\,eV/\AA, which is not as small as the most recent MLIP models. This can be attributed to (1) the employed high temperature, (2) the out-of-equilibrium nature of the growth process and (3) the hybrid nature of PLIP+Q which is positioned in between purely ML and classical IP.

\section{Influence of the deposition rate}

\begin{figure}[h!]
    \centering
    \includegraphics[width=17cm]{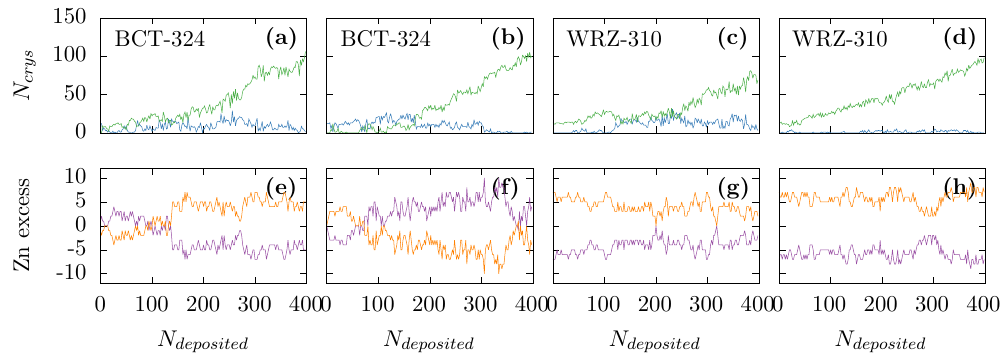}
    \caption{Results obtained at a slower deposition rate. (a) to (d): Examples of the evolution of the number of atoms of the BCT (blue) and WRZ (green) type inside the crystalline nucleus as a function of the number of deposited atoms obtained for BCT-324 and WRZ-310 initial seeds at 500 K. (e) to (h): Corresponding excess of Zn atoms at opposite ends of the nanoparticle, in orange and purple, respectively.}
    \label{Slow}
\end{figure}

To confirm our results, we carried out additional simulations with the deposition rate reduced by a factor of five which means a deposition every 50\,ps. Fig.\,\ref{Slow} shows typical crystallization and Zn excess curves obtained for the WRZ-310 and the BCT-324 systems. Results using this slower deposition rate are very similar both qualitatively and quantitatively to the previous results.

\section{Surface mobility of deposited atoms}

\begin{figure}
    \centering
    \includegraphics{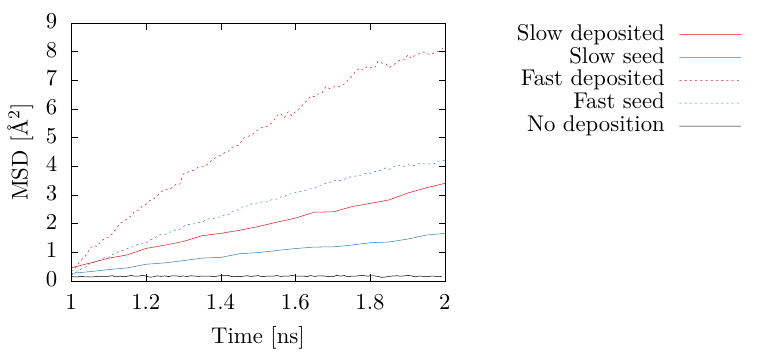}
    \caption{Mean square displacement averaged over all simulations for the BCT-162 seed at two different deposition rates: one ZnO pair deposited every 50 ps (solid lines) and every 10 ps (dashed lines). }
    \label{Diffusion}
\end{figure}

A common challenge in growth modeling is that certain simulation parameters may not allow for sufficient mobility of the deposited atoms, potentially causing the system to remain trapped in irregular morphologies. To assess whether this occurred in our simulations, we computed the mean square displacements using two different deposition rates and compared these results with simulations performed without any deposition. As shown in Fig.\,\ref{Diffusion}, we find that (1) the deposition process generally enhances atomic mobility, (2) slower deposition rate leads to lower mobility, and (3) the deposition induces additional displacements even for atoms belonging to the original seed.

\section{Convergence of the mean}

\begin{figure}[ht]
\centering
    \includegraphics[width = 17cm]{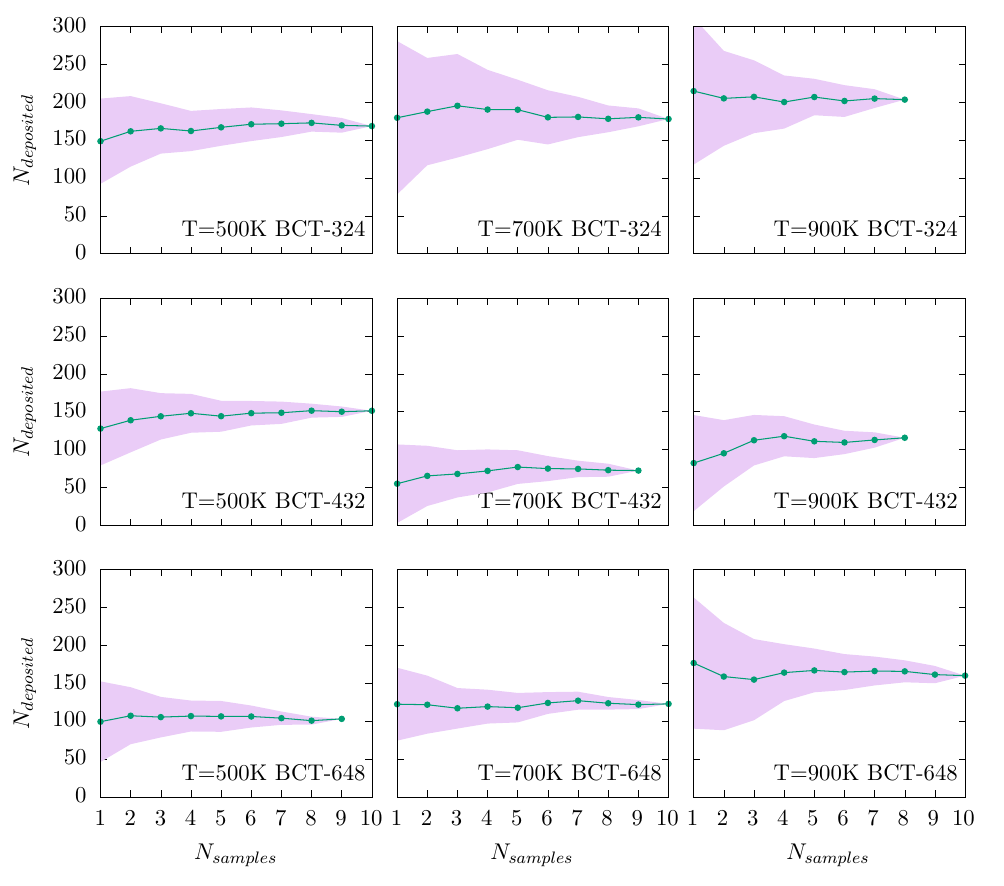}
    \caption{Convergence of the mean results with in green the obtained subsamppled mean and in violet the corresponding standard deviation. We note that since the BCT$\rightarrow$WRZ transition does not occur in all the cases, some convergence of the mean calculations were done with less than 10 samples.}
    \label{Stat}
\end{figure}

To analyze if 10 simulations are sufficient, we performed the convergence of the mean which is a more thorough stability analysis. In particular, for values of $N_{samples}$ going from 1 to 9, we randomly selected subsets of $N_{samples}$ of our 10 measurements and measured the corresponding average. By doing this randomization 20 times, we can obtain a mean and a standard deviation of the average when using only $N_{samples}$ instead of the 10 measurements. For each values of the temperature and the initial seed size, Fig.\,\ref{Stat} shows that both mean and the standard deviation of the average starts by largely fluctuating when using only 1 to 5 samples but then it reaches a plateau starting from 7 until 10 samples. Quantitatively, this results at most to an error of 9\% when using 9 instead of 10 simulations. This confirms that 10 simulations should be sufficient to provide a quantitative measurement of the number of deposited atoms required to reach the BCT$\rightarrow$WRZ transition.

\section{Delayed WRZ growth}

\begin{figure}[h!]
    \centering
    \includegraphics{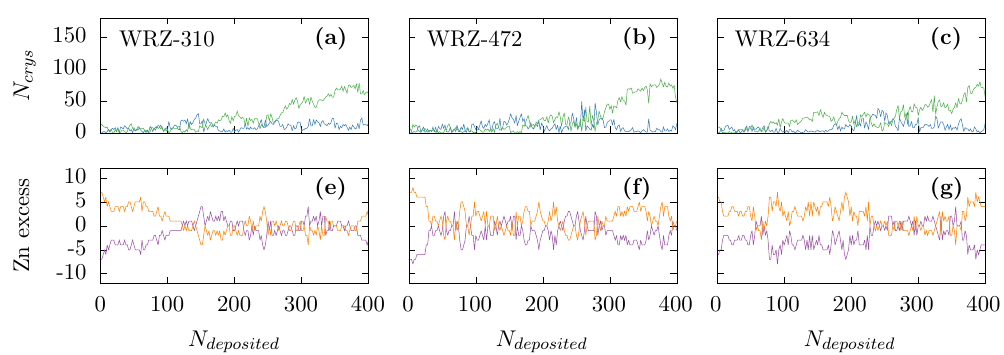}
    \caption{Results for the cases in which growth of WRZ phase is substantially delayed. (a) to (c): Evolution of the number of atoms of the BCT (blue) and WRZ (green) type inside the crystalline nucleus as a function of the number of deposited atoms obtained for WRZ-310, WRZ-472 and WRZ-634 initial seeds. (d) to (g): Corresponding excess of Zn atoms at opposite ends of the nanoparticle, in orange and purple, respectively.}
    \label{QuenchWRZ}
\end{figure}

Fig.\,\ref{QuenchWRZ} shows the few cases where we observe a delayed growth of WRZ when starting from WRZ seed. In all cases, we observe a concomitant quench of the Zn excess.

